# Plasma Heating by an Electron Beam in Corrugated Magnetic Field

V.V. Postupaev, A.V. Arzhannikov, V.T. Astrelin, A.M. Averkov, A.D. Beklemishev, A.V. Burdakov, I.A. Ivanov, M.V. Ivantsivsky, V.S. Koidan, K.I. Mekler, S.V. Polosatkin, A.F. Rovenskikh, S.L. Sinitsky, Yu.S. Sulyaev, E.R. Zubairov

*Budker Institute of Nuclear Physics, Novosibirsk 630090, Russia*

New experimental results from the multimirror trap GOL-3 are presented. Deuterium plasma of ~$10^{21}$ $m^{-3}$ density is heated by a high power relativistic electron beam (peak parameters are ~1 MeV, ~25 kA, ~8 μs, ~120 kJ). Magnetic system of the facility is a 12-meter-long axisymmetrical solenoid with corrugated magnetic field, which consists of 55 cells with $B_{max}/B_{min}$=4.8/3.2 T. Collective plasma heating by the electron beam results in peak electron temperature ~2 keV. To this time the ions are also collectively heated by gradients of electron pressure in the cells of the trap. Ion temperature increases above 1 keV and confines at the high level for ~1 ms. Dense hot plasma in GOL-3 trap emits D-D neutrons for ~1 ms.

Details of collective plasma heating by the beam in the corrugated magnetic field will be presented in the paper. New physical mechanism of effective heating of plasma ions, substantially dependent on the corrugation of the magnetic field, is discussed. Experiments with complete multimirror configuration of the GOL-3 facility have shown the significant improvement of energy confinement time comparing with simple solenoidal configuration. Axial currents, which exist in the system, cause sheared helical structure of the magnetic field. The safety factor $q$ is shown to be below unity on the axis.

## Introduction

Multimirror confinement of hot dense plasma in a long open trap is studied at GOL-3 facility in Novosibirsk [1]. The multimirror trap consists of a set of mirrors, which are connected to each other at their ends. Full solenoid length $L$ exceeds mean free path of the particles λ. If, at the same time, the mirror cell length $l<\lambda$ then the longitudinal confinement time increases essentially compared to classical single mirror trap due to diffusion-like plasma expansion (see review [2]).

This paper covers our new experiments with multimirror magnetic system at GOL-3 facility with the main focus on features of the plasma heating. The plasma is heated by an electron beam up to keV-range temperatures due to non-classical, collective processes. Plasma stability and confinement depends on the structure of the magnetic field, which is also formed under conditions of anomalous collisionality.

The current of the injected electron beam exceeds significantly the Kruskal-Shafranov limit for kink stability. This problem of stable transportation of the beam in the long plasma system was solved by preparation of adequate preliminary plasma with high enough compensating return current. In the case of a significant decay of the return current the helicity of the magnetic field changes sign and the safety factor $q$ falls below unity in the plasma core. Measurements suggest that this is the case at least during the beam phase of the discharge.

## GOL-3 Facility

GOL-3 facility is a long open trap intended for study of heating and confinement of a relatively dense ($10^{21}$÷$10^{23}$ $m^{-3}$) plasma in axisymmetrical magnetic system [1]. The magnetic system consists of coils for transport and shaping of the electron beam, a 12-meter-long solenoid and an exit unit (which includes plasma creation system, expander and exit receiver of the beam and plasma). The solenoid consists of 55 cells of 22 cm length each with $B_{max}/B_{min}$= 4.8/3.2 T (Fig.1). Preliminary deuterium plasma of $(0.2\div5)\cdot10^{21}$ m$^{-3}$ density is created by a special linear discharge along the whole device length. Axial distribution of the plasma density is formed by a set of fast gas-puff valves. The plasma heating is provided by a high-power electron beam (~1 MeV, 30 kA, 8 μs) with total energy content of 120÷150 kJ.

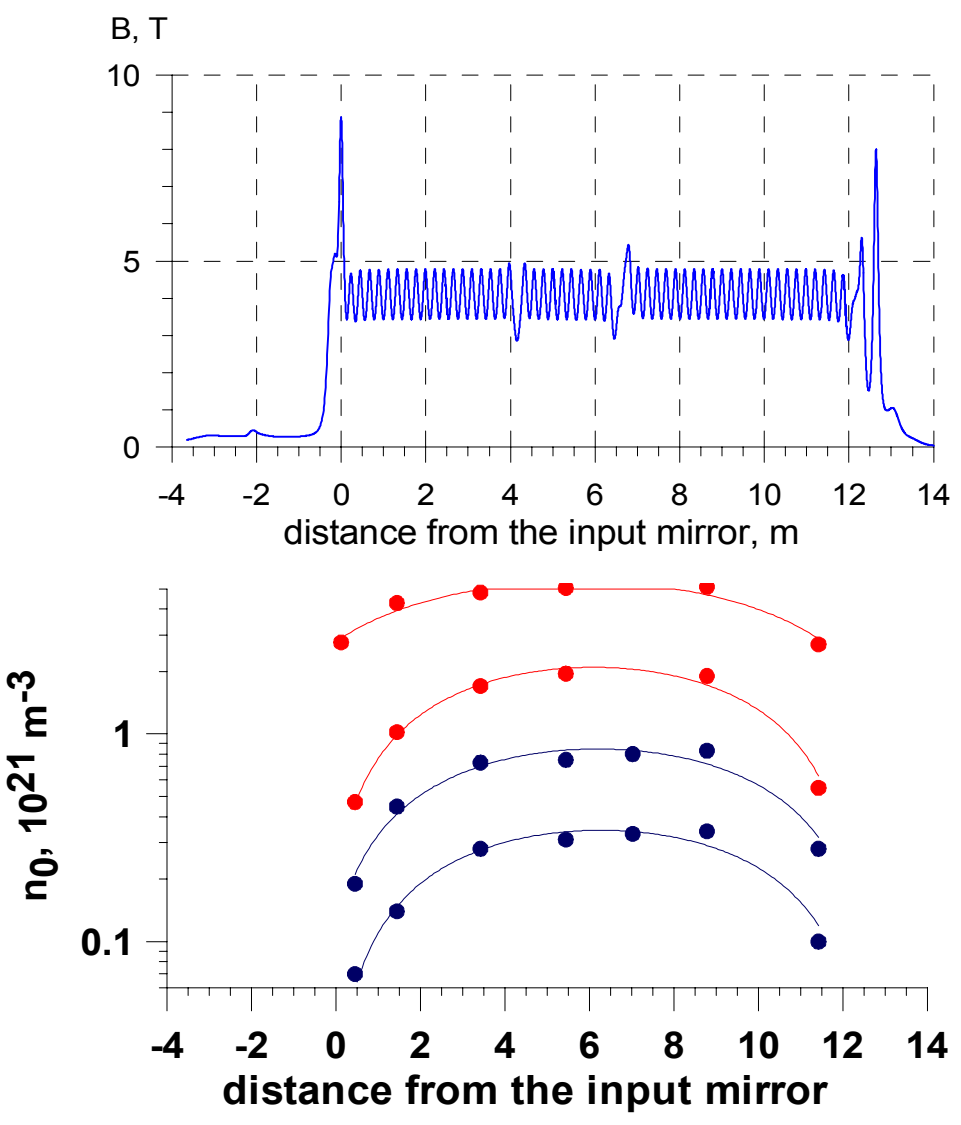

*Fig.1. Magnetic field configuration (top). Axial distribution of the initial deuterium density $n_0$ (bottom). Four different regimes are shown.*

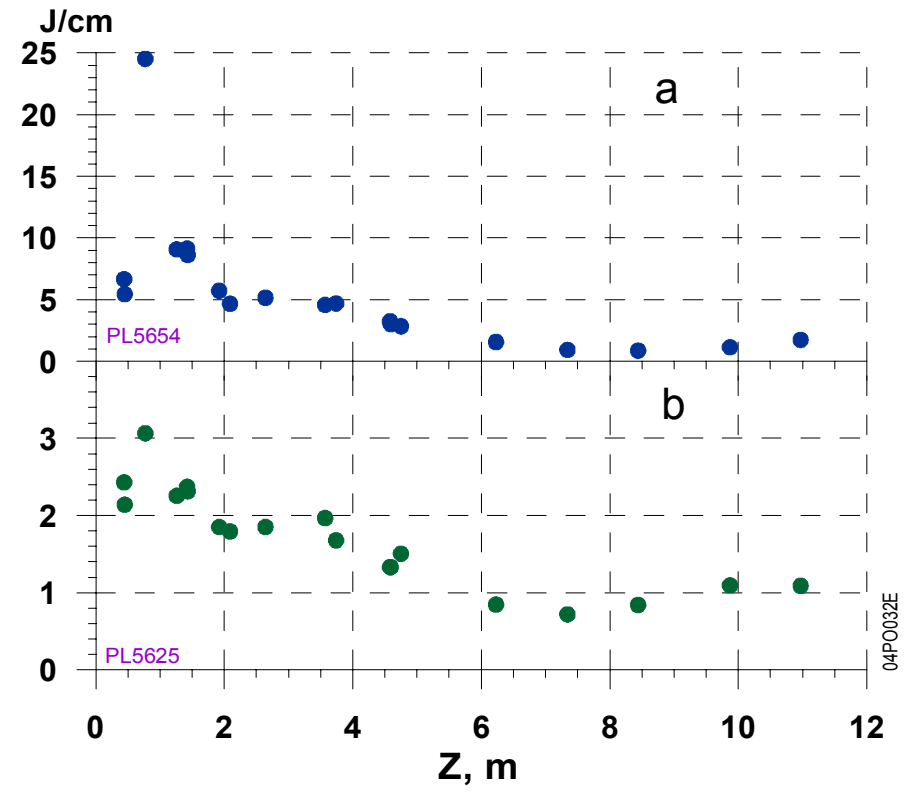

*Fig.2. Axial distribution of plasma energy (multiplied to local cross section) at different initial density.*

## Plasma Heating Stage

The relativistic electron beam interacts collectively with the plasma and excites a high level turbulence in it. As a result of collective interactions, the electron distribution function becomes essentially a nonequilibrium one. The bulk of electrons is heated up to average energies (temperature) of 1÷2 keV, the tail extends up to the beam energy [3]. In this case the spatial anisotropy of the electron distribution function is observed: the "axial" mean energy of electrons exceeds by several times the "radial" one. The suprathermal electrons keep the major part of energy left by the beam in the plasma. The beam energy release is nonuniform along the system length. Fig.2 shows the axial distribution of specific plasma energy for two operation regimes. Dependence *a*) corresponds to a high temperature regime with rather low density $(0.2\div0.5)\cdot10^{21}$ m$^{-3}$, *b*) corresponds to a high density regime $(4\div5)\cdot10^{21}$ m$^{-3}$. It is seen that in the high-$T$ regime there is the pressure peak at a distance of about 1 m from the entrance mirror. In the high-$n$ regime this dependence is smoother.

As a result of the beam-plasma collective interaction the effective collision frequency of plasma electrons exceeds by three orders the classical Coulomb collision frequency. Consequences of this [4] are that the plasma electrical conductivity and its longitudinal electron thermal conductivity decreased substantially. All the mentioned above phenomena enable a possibility to form the high pressure gradients along the magnetic field and macroscopic motion of a plasma. As a result, fast heating of ions is observed in GOL-3 [5].

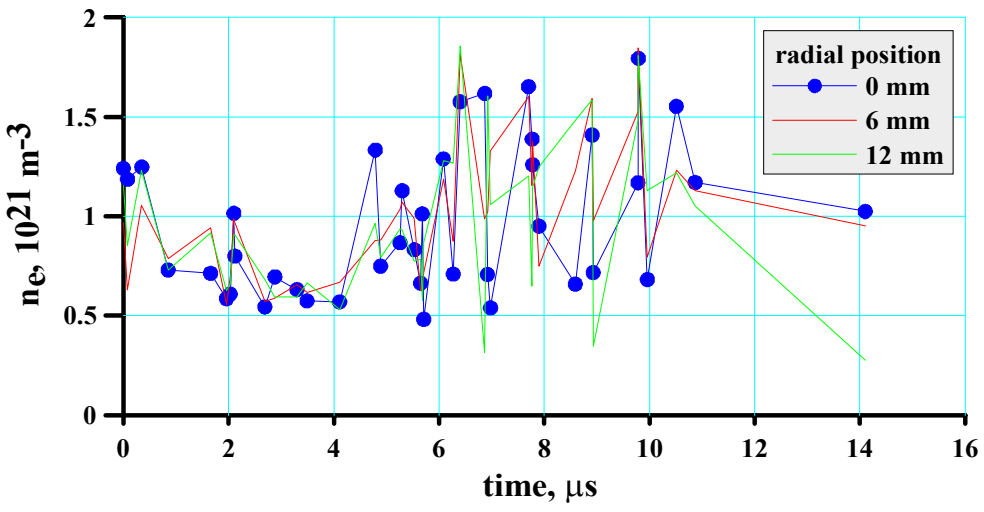

*Fig.3. Shot-by-shot Thomson data for $n_e(r)$.*

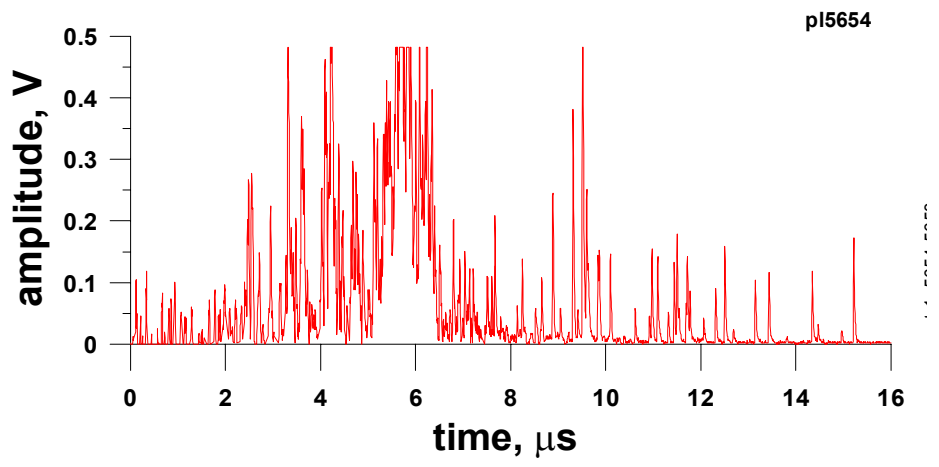

*Fig.4. Initial phase of neutron emission, $n_0=0.3\cdot10^{21}m^{-3}$.*

## Fast Heating of Ions (Model and Experiment)

A model for explanation of fast ion heating in a multimirror trap was proposed (see [5]). It takes into account the following: a) a nonuniform plasma heating (which depends on the $n_b/n_p$ ratio, i.e. on the local magnetic field); b) suppression of heat transport during the heating phase that enables high pressure gradients; c) collective acceleration of plasma flows from the high-field part of corrugation cells to cell's 'bottom'; d) thermalization of the opposite ion flows. The model predicts occurrence of strong modulation of density and velocities of ions in the trap after several microseconds of beam injection. This mechanism of ion heating was studied with a Thomson scattering system (10 J, 1.06 μm), which was used to measure an 8-point $n_e(r)$ distribution. Shot-to-shot spread of the plasma density is below 10% during several first microseconds of the beam injection. Then the density spread becomes larger, up to $\delta n/n$~40% (Fig.3). Measurements of a neutron emission intensity (Fig.4) show that first neutrons are detected in 2÷3 μs after beginning of the beam injection and then a powerful flash of neutron radiation takes place in 4÷8 μs. Axial distribution of the pressure (Fig.2) in general corresponds to distribution of neutron emission along the trap length. There is a pressure peak at a distance of about 1 m from the input mirror. At this point a localized maximum of the neutron emission is observed. The plasma motion in the axial direction and redistribution of its parameters along the radius is accompanied by excitation of oscillations inside the hot part of the plasma column. This is clearly detected by neutron and some other local diagnostics.

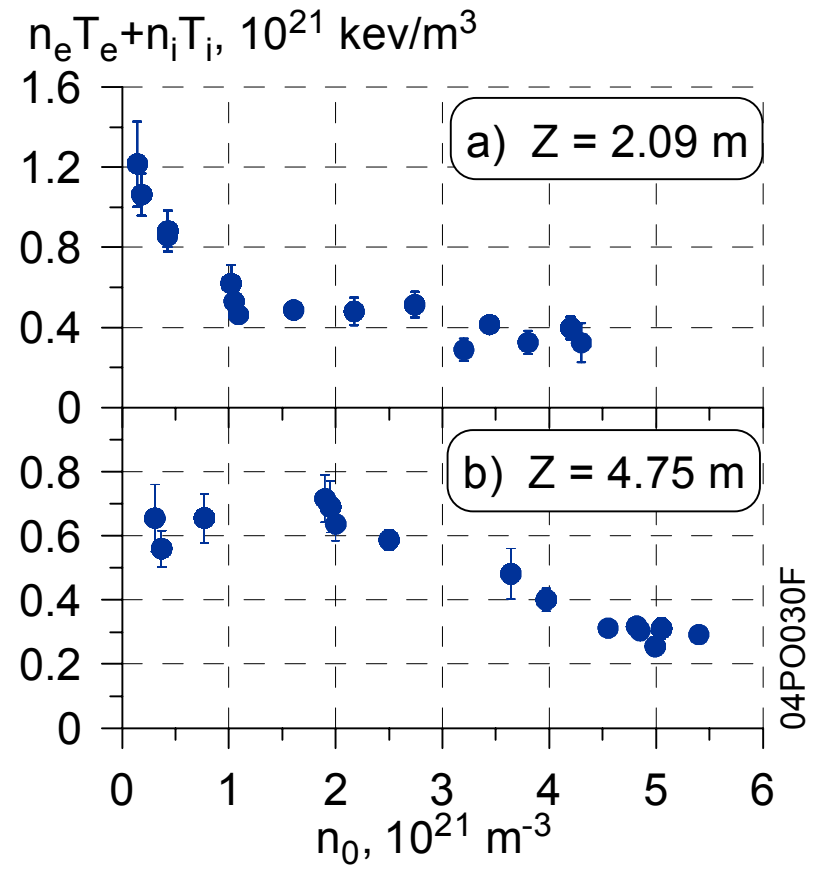

*Fig.5. Plasma pressure vs. initial deuterium density for t=10 μs.*

## Dependence of Plasma Parameters on Density

Dependence of a "diamagnetic" plasma pressure on the initial deuterium density is shown in Fig.5 for two distances from the input mirror. The time moment of 10 μs after the beginning of the beam injection characterizes the plasma pressure after completion of the fast heating of electrons and ions. The beam ends 2÷3 μs earlier and part of supra-thermal electrons leaves the trap to this time. At the device entrance, the dependence is strong but in the central section of the trap the plasma pressure is practically independent on density up to ~$2\cdot10^{21}$ m$^{-3}$. After that, it drops slowly. Wall calorimeter data shows that in the high-$T$ regime, the transverse losses (mainly for radiation) remain to be lower than longitudinal ones. At the high-$n$ regime the fraction of transverse losses increases and moreover, the losses to limiters become probable.

## Currents in GOL-3 Plasma

An important feature of GOL-3 facility is that it has two external sources of axial current. The first one is the plasma source, which creates preliminary (target) low-temperature plasma. by a linear discharge through the entire solenoid length. The second is the U-2 generator of the relativistic electron beam, which injects the beam into the preliminary plasma. The net plasma current is almost unchanged during the beam injection, it means that the return current, which is opposite to the beam current, is generated in the plasma. The discharge current achieves 7 kA at low density and decreases to ~1 kA at high density. As shown earlier (see, e.g., [6]), when the discharge current falls below some threshold value, compensation of the beam current becomes worse, full current exceeds the Kruskal-Shafranov limit and the beam propagation through the plasma becomes unstable.

The current density on the axis is almost equal to the averaged beam current density and it is co-directed with the beam current [7]. Comparison of the net plasma current (which is counter-directed to the beam current) with the current on the axis (which is co-directed) shows that there is a surface with $B_\phi=0$ inside the plasma, and the safety factor

$$q = \frac{2\pi r B_z}{L B_\phi}$$

goes to infinity at this surface (here $L$ and $r$ are the plasma length and radius, $B_z$ and $B_\phi$ are the components of the magnetic field).

## Direct Measurement of *q* Profile

A new specialised diagnostic was developed for direct measurement of the helicity of the field lines during the beam injection period. For this purpose a fast-frame pinhole detector imaged the X-ray footprint of the beam on the exit collector. The narrow graphite bar was placed into the beam cross-section near the far end of the plasma source. This bar partly blocks the electron beam and creates an "electron shadow". The beam electrons, after passing the bar position, move essentially along the helical field lines. Rotation of the magnetic field was measured by the print of the beam shadow on the collector. Shear of the magnetic field, resulting from the partial compensation of the beam current, is observed.

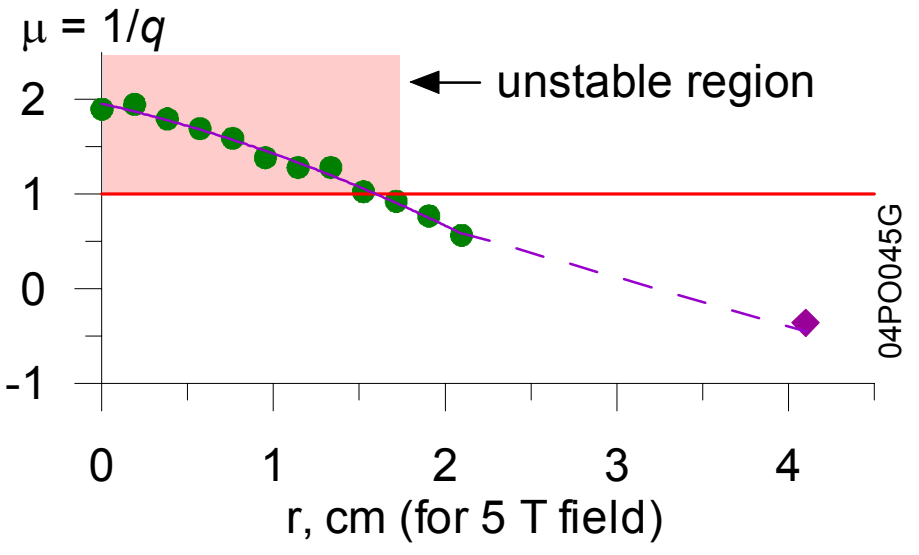

*Fig.6. Typical radial dependence of μ=1/q. Dots are calculated from the X-ray image, diamond is calculated from the net current.*

The measured dependence *q(r)* is shown in Fig.6. We notice a large paraxial zone with $q<1$, i.e., unstable with respect to the $m=1$ mode. Possible indications of such instability are the X-ray images made with larger delays from the beam start, some of which have much wider (or even absent) beam shadow. Such events mean that during the frame duration of 1 μs a global reorganization of the magnetic structure occurs at least within the beam cross-section. The graphite bar partly blocks the current, so that the paraxial μ value is a floor estimate.

## Influence of *q* Profile on Heating

Stabilisation by shear is an important factor, which affects the plasma confinement. The discharge current oscillates and we can change delay between the discharge and the beam. A special experiment with different delays was performed (all other parameters were fixed). Fast full ionisation occurs after the beam start. This makes initial conditions identical except for different amplitudes of the plasma current. Typical waveforms are shown in Fig.7. Delay of 45 μs is the main one for GOL-3 facility because it provides stable operation in a widest density interval. Changing delay of the beam start, we can perform shots with the discharge current co- or counter-directed to the beam current, or with the discharge already completed. The degree of decompensation depends on the delay. From the second half-period and later the current can reach dangerous global *q* values, at which all the plasma becomes unstable with respect to the *m*=1 kink mode.

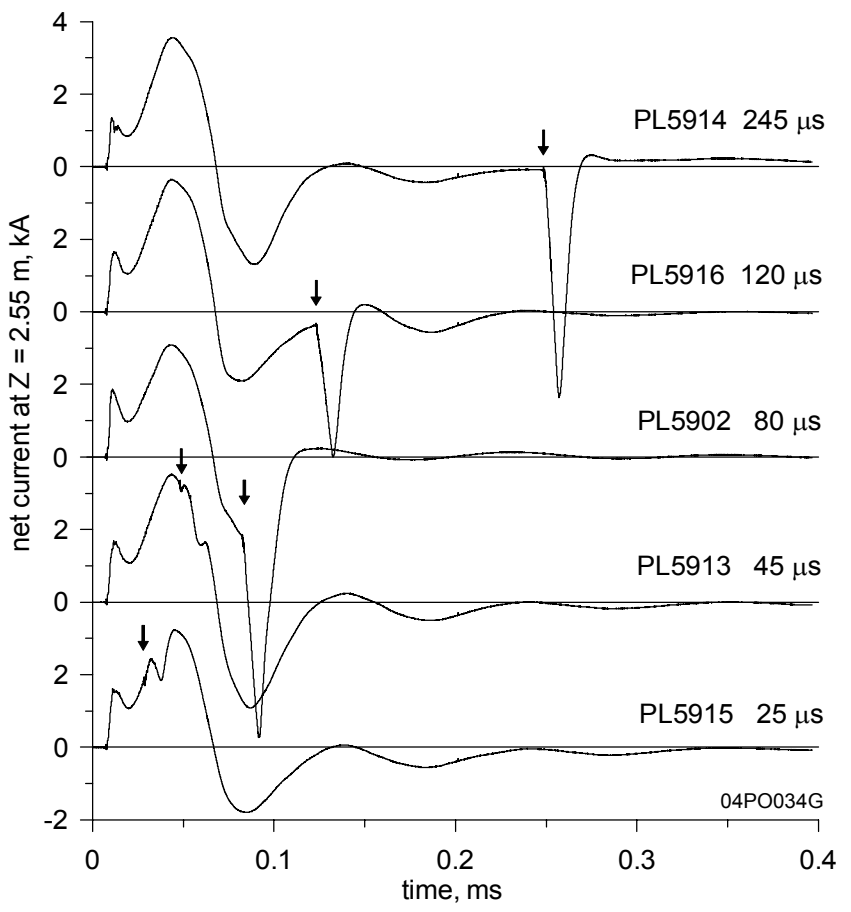

*Fig.7. Net plasma current at different delays of the beam injection.*

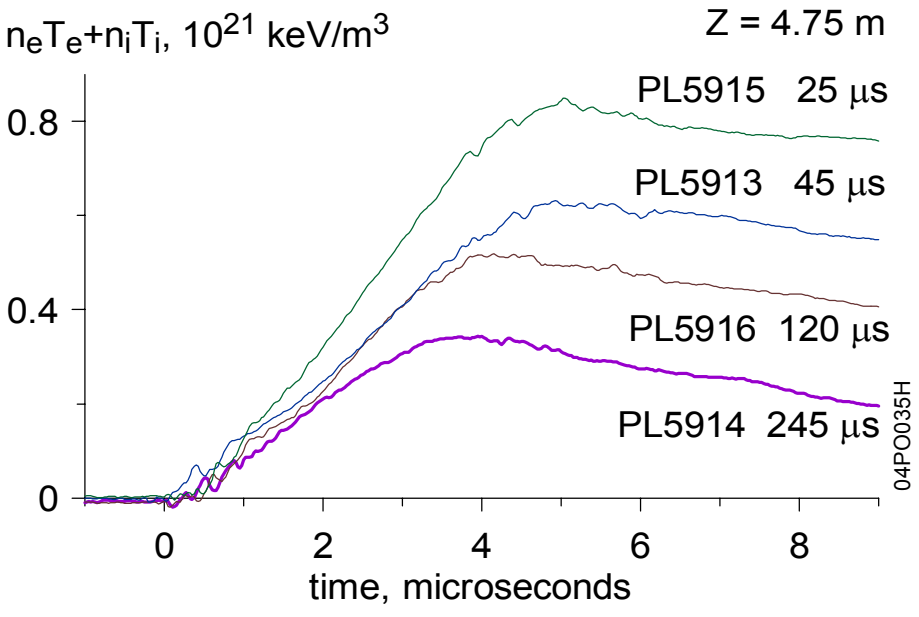

*Fig.8. Dependence of the plasma diamagnetism on the delay of the beam injection (t=0 is the beam start).*

Typical diamagnetic signals for *Z*=4.75 m are shown in Fig.8. Some conclusions from this series of shots are:

1) Change of the net current depends on the delay (this gives initial and final *q* profiles).

2) Peak plasma pressure and period of pressure growth depend on the delay.

3) At the first few meters of the plasma (region with highest beam-plasma interaction) there is an optimum delay time.

4) At farther sections of the plasma column (e.g. as in Fig.8) the peak value of the pressure monotonously decreases with the increase of the delay.

Results 2-4 are really unexpected. A possible explanation is that if a surface with $B_\phi$=0 exists inside the plasma, then the structure of turbulence changes and the level of turbulence somehow affects the beam-plasma interaction. The experiment performed shows that the radial structure of currents is really a factor, which determines the plasma confinement also.

## Discussion

Unlike most open traps, GOL-3 uses longitudinal beam current for heating purposes. As a result, its magnetic field has helical structure with its inherent advantages and drawbacks. The advantage is the stabilizing influence of the magnetic shear on the interchange modes. On the other hand, the azimuthal magnetic field carries a significant free energy and thus can act as a source of additional instabilities. Such are kink and tearing modes, which are absent in usual mirror traps but are common in tokamaks.

Indeed, in some ways the field structure and stability properties of GOL-3 are similar to those in tokamaks. Most notable similarities are 1) quasi-periodic modulation of curvature along the field line, 2) helical magnetic field, 3) one degree of axial symmetry (which is poloidal rather than toroidal, though). Meanwhile, the differences can be listed as 1) the absence of true periodicity, "openness" of the trap, 2) the role of poloidal field modulation is taken over by multi-mirror modulation, which corresponds to tokamak "ripples" rather than toroidicity.

Our current understanding of the field formation and evolution is as follows:

1) Before the beam-injection phase, we form a preliminary Ohmic discharge with a significant counter-beam current. This is found to be necessary in order to reduce the average azimuthal field and avoid violent external kink instability later.
2) During the heating phase the effective electrons collision frequency increases by a factor of $\sim 10^3$ due to turbulence [4]. Though this period lasts just 5-10 μs, the poloidal magnetic field of the beam penetrates the plasma column, so that the on-axis value of the safety factor $q$ reverses its sign (as compared to the preliminary discharge) and falls below unity.
3) Direct measurements show $q(0)\sim 0.5$ during the beam phase. Indeed, one would expect development of the $m=1$ internal tearing mode (sawtooth) when $q(0)$ falls below unity.

The boundary conditions for the tearing mode in our case are not periodic, which changes the threshold to ~0.8. Kadomtsev's estimate for the reconnection time gets a time-scale which is shorter than the beam-injection phase. Thus, though such sawteeth cannot be at present clearly distinguished, it is reasonable to expect reconnection events developing during the beam phase in the core of the discharge.

Observed role of the magnetic shear is extremely important. If we consider a multimirror trap as a reactor, then the magnetic shear can provide stability at $\beta<1$. Existing theory (see, e.g., [2,8-9]) does not cover all aspects of the only one existing multimirror experiment, GOL-3, with large longitudinal currents. Therefore, the theory considers only plasma with $\beta>1$ to be stable (so-called "wall confinement") with heavy doubts on the possibility of achievement of such regime and safe transition through the unstable $\beta<1$ region. Stabilizing role of the magnetic shear enables us to propose GOL-3-like systems for studies of plasma with advanced sub-reactor parameters.

## Conclusions

Main results of this work are the following:

1. Conditions for effective plasma heating in the range $(0.2\div6)\cdot10^{21}$ m$^{-3}$ are experimentally found. Electron and ion plasma temperatures up to 2 keV at density $\sim10^{21}$ m$^{-3}$ are achieved.
2. Direct measurements of $q$ profile were done. Radial profile of the return current was shown to differ from that of the beam current, i.e. there is no full compensation by current density.
3. Radial structure of currents results in a sheared magnetic field, which can stabilize some MHD modes in the multimirror trap. In the paraxial area $q < 1$ during the beam injection, this area is unstable with respect to the internal $m$=1 tearing mode.
4. Experiments with varying delay of the beam injection show that heating and confinement of plasma in GOL-3 substantially depend on the $q$ profile.

## Acknowledgements

Authors are grateful to other members of GOL-3 team for joint work in the experiments. This work is partially supported by Russian Ministry of Education and Science and by RFBR projects 03-02-16271a and 04-01-00244a.